\documentclass[twocolumn,pra,showpacs]{revtex4}
\usepackage{bm}
\usepackage{graphicx}

\begin{document}

\title{Reply to "Comment on 'Detecting Non-Abelian Geometric Phases with Three-Level $\Lambda$ Atoms' "}
\author{Yan-Xiong Du}
\author{Zheng-Yuan Xue}
\author{Xin-Ding Zhang}
\author{Hui Yan}
\email{yanhui@scnu.edu.cn}

\affiliation{Laboratory of Quantum Information Technology, School of
Physics and Telecommunication Engineering, South China Normal
University, Guangzhou 510006, China}

\begin{abstract}
In this reply, we address the comment by Ericsson and Sjoqvist on
our paper [Phys. Rev. A {\bf 84}, 034103 (2011)]. We point out
that the zero gauge field is not the evidence of trivial geometric
phase for a non-Abelian SU(2) gauge field. Furthermore, the
recalculation shows that the non-Abelian geometric phase we
proposed in the three-level $\Lambda$ system is indeed
experimentally detectable.
\end{abstract}
 \maketitle

There are three points in the preceding comment \cite{Ericsson}:
(i) the sign in Eq.(7) in the original paper\cite{du} is wrong;
(ii) non-Abelian geometrical phase(GP) of a large-detuned
$\Lambda$ system ( $\sin \gamma\approx
 0 $ with $\gamma$ defined below) in the adiabatic approximation is
trivial since the corresponding gauge field vanishes; (iii) the GP
is still small in the general case with relatively large $\gamma$
and cannot be separated from the dynamical phase .

In this reply, we address the above comments. We agree with point
(i). However, this mistake doesn't influence the validity of the
main result. We re-calculate the population difference induced by
the geometric phase after correcting this sign error and find that
the maximum population difference can still reach $20\%$. Such big
population difference can be easily detected in a typical
experiment.

We focus on the large-detuning case, where
$\tan\gamma=(\sqrt{\Delta^2+\Omega^2}-\Delta)/\Omega \approx 0$ with
$\Delta$ $(\Omega)$ the detuning (effective Rabi frequency). The
large detuning can be easily realized through choosing proper parameters, i.e.
$\Delta=1$ GHz and $\Omega=1$ MHz. As corrected in the comment\cite{Ericsson}, the gauge potentials in the
large detuning case can be written as
\begin{eqnarray}
A_\theta& \approx& i\cos{\varphi}\sigma_y+i\sin{\varphi}\sigma_x,\\
A_\varphi&\approx & -i\sin^2{\theta}\sigma_z+i\sin{\theta}\cos{\theta}\cos{\varphi}\sigma_x\\
\nonumber &&-i\sin{\theta}\cos{\theta}\sin{\varphi}\sigma_y.
\end{eqnarray}
Then we also agree that the corresponding gauge field
$F_{\theta\varphi}$ must vanish.

One will think of such potentials like Eq.(1) and (2) may not
bring any physical observable effects because of the vanished
field strength. In the following we will illustrate with some
examples that this granted judgement is wrong for non-Abelian
gauge fields. The first example is about the vacuum solution in a
Yang-Mills theory demonstrated in detail in Ref.\cite{Jackiw}.
Jackiw and Rebbi demonstrated that\cite{Jackiw}, a type of gauge
potentials in the Yang-Mills theory are gauge equivalent to the
potential $A=0$ and thus the field strengths are vanished;
however, the potentials should not be removed from the
integrations over the field configurations by the gauge fixing
procedure, and they argued that physical effects are associated
with them. The second example we will address in detail is the
well-known case that spin $1/2$ behavior in a time-dependent
magnetic field $\mathbf{B}=(B_x(t), B_y(t),
B_z(t))$\cite{Berry,Bliokh}. The interacting Hamiltonian is
\begin{equation}
H=\hbar\chi\mathbf{\sigma}\mathbf{B},
\end{equation}
here $\sigma$ is the Pauli Matrixes and $\chi$ is the coupling constant.
This Hamiltonian can be diagonalized through $H_{dg}=U^{-1}HU$ with the unitary matrix
\begin{equation}
U=\frac{1}{\sqrt{2B}}\left(\begin{array}{cc}
\sqrt{B+B_z} & \sqrt{B-B_z}\\
\frac{B_x+iB_y}{\sqrt{B+B_z}} & -\frac{B_x+iB_y}{\sqrt{B-B_z}}\\
\end{array}\right),
\end{equation}
here the corresponding eigenvalues are $E_{\pm}=\pm\hbar\chi B$,
$B=|\mathbf{B}|$. For a pure gauge potential with the unitary matrix
(4) as
\begin{equation}
\mathbf{A}(\mathbf{m})=iU^{-1}\frac{\partial U}{\partial\mathbf{m}},
\mathbf{m}=(B_x, B_y, B_z),
\end{equation}
the non-Abelian gauge field $\mathbf{F}$ is vanishing \cite{Bliokh}.
To recover the non-trivialness of such potential, one should choose
a specific closed loop in this parameter space, i.e. , a loop with
large enough $B$ to maintain the adiabatic approximation. In this
case, the off-diagonal elements of $\mathbf{A}(\mathbf{m})$ are
rejected and the adiabatic gauge potentials $\mathbf{A}^{(ad)}$ are
given by
\begin{equation}
\mathbf{A}^{(ad)}=(\frac{B_y}{2B(B\pm B_z)},-\frac{B_x}{2B(B\pm
B_z)},0),
\end{equation}
where different signs correspond to two eigenvalue $E_{\pm}$. The
corresponding field tensors are given by
\begin{equation}
F_{ij}^{(ad)}=\frac{\partial A_j^{(ad)}}{\partial
m_i}-\frac{\partial A_i^{(ad)}}{\partial
m_j}=\mp\varepsilon_{ijk}\frac{B_k}{2B^3},
\end{equation}
$\varepsilon_{ijk}$ is the three order antisymmetric tensor. It
can be seen that the closed path integral $\oint
\mathbf{A}^{(ad)}d\mathbf{l}$ in these monopole field strength is
just the case of Berry phase. This integral is non-zero and will
bring physical observable effects. Furthermore, the integral
should hold when we shrink the curve by decreasing $B$ linearly,
then we return to the case of full matrix
$\mathbf{A}(\mathbf{m})$. Therefore, $\oint
\mathbf{A}(\mathbf{m})d\mathbf{l}$ also contains a flux and will
bring physical observable effects. Namely, the non-vanishing
$\oint \mathbf{A}(\mathbf{m})d\mathbf{l}$ is guaranteed by $\oint
\mathbf{A}^{(ad)}d\mathbf{l}$ in the Abelian case. And hence, we
show with the second exampe that the gauge potential with a zero
gauge field will also have observable effects for its
non-vanishing closed path integral in the non-Abelian case.

\begin{figure}[ptb]
\begin{center}
\includegraphics[width=7cm]{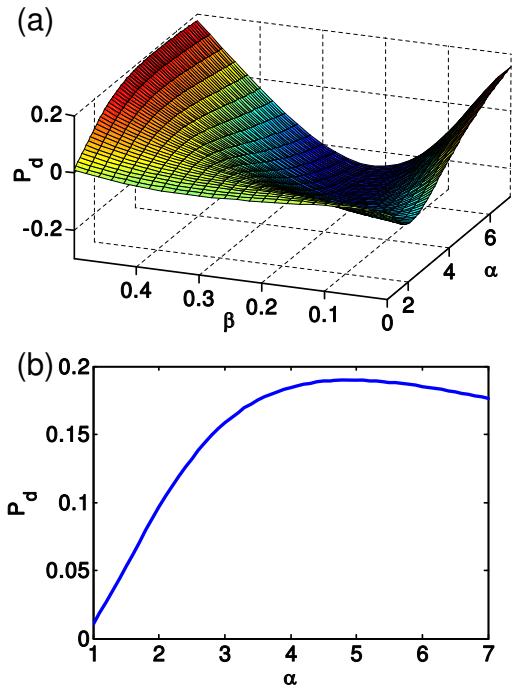}
\caption{\label{fig:1}(color online).(a) The new result of
population difference after two composed path with different orders.
(b) The population difference $P_d$ versus parameter $\alpha$ when
$\beta=0.5$. The parameters $\alpha$ and $\beta$ represent the
relative amplitude and the time delay of Raman pulses in the
stimulated Raman adiabatic transitions as the same in the origin
paper \cite{du}.}
\end{center}
\end{figure}

The situation of three-level $\Lambda$ atoms interacting with lasers
is similar with the above analyzation. The unitary matrix
diagonalize Hamiltonian (1) in \cite{du} reads as
\begin{equation}
\Gamma=\left(\begin{array}{ccc}
\cos{\theta}&-\sin{\theta}e^{-i\varphi}&0\\
\sin{\theta}\cos{\gamma}e^{i\varphi}&\cos{\theta}\cos{}\gamma&-\sin{\gamma}\\
\sin{\theta}\sin{\gamma}e^{i\varphi}&\cos{\theta}\sin{\gamma}&\cos{\gamma}
\end{array}\right),
\end{equation}
and $\gamma$ is given by
$\tan{\gamma}=(\sqrt{\Delta^2+\Omega^2}-\Delta)/\Omega$. One can set
large detuning $\Delta$ to achieve the non-Abelian geometric phase
Eqs. (1,2). The corresponding gauge field is zero. To recover the
non-trivialness of Eqs. (1,2), suitable $\Delta$ and $\Omega$ should
be chose to reject the non-diagonal elements. Then we return to the
Abelian case of which the gauge potential is $A^\pm_\varphi= \mp
\sin^2{\theta}$ is observable \cite{Zhu}. The non-zero closed path
integral of $A^\pm_\varphi$ can be extended to the non-Abelian case
just as the above analyzation. For example, we choose a closed loop
of $\mathbf{A}=(A'_\theta, A'_\varphi)=i(A_\theta, A_\varphi)$ along
$C$ with $\theta=\theta_0$, $\varphi\in [0,2\pi]$, and then the
integral can be derived as
\begin{equation}
\oint \mathbf{A} d\mathbf{R}=\int_0^{2\pi}A'_\varphi d\varphi
=\pi(1-\cos2\theta_0)\sigma_z=\frac{1}{2}\Omega(C)\sigma_z,
\end{equation}
where $\Omega(C)$ is the solid angle spanned by $C$ which shows the
geometric feature of the evolution.  Clearly this integral is
generally non-vanishing and then the corresponding  geometric phase
is non-trivial.

The result of a vanishing gauge field brings physical effects is
counterintuitive. Indeed, this comes from the fact that the Stroke
theorem in the non-Abelian case is not a direct generalization of
the Abelian case. It has been realized that the surface integral in
the non-Abelian case should depend on the gauge field
$F_{\theta\varphi}$ as well as the gauge potential
$\mathbf{A}$\cite{niu,Barrett}. Actually, the phase factor given as
\begin{equation}
\label{Wilson}
 U= \hat{P} \exp \left(i \oint \mathbf{A}\cdot d\mathbf{l}\right)
 \end{equation}
with $\hat{P}$ being the operator of chronological ordering, which
is called the Wilson loop\cite{Wilson}, is of particular important
for a non-Abelian field. This phase factor is transformed
covariantly and may results in the observable physical effects even
with zero field strength\cite{Bliokh,Yang}.

We re-calculate the possibly observed effects based on Eqs.(1) and
(2).  The new results are show in Fig.1, which should replace the
results of Fig.2 in \cite{du}. The  procedure and parameters are
the same with those in our original paper\cite{du}, but the gauge
potentials are replaced with those in Eqs(1) and (2). We can find
from Fig.1 that the maximum population difference can still reach
almost $20\%$. Such big population difference can be easily
detected in a typical experiment. Since $\gamma \approx 0.001$ for
$\Gamma/\Delta=0.001$, we know that the effects induced by the
dynamical phase could be neglected in the above calculation, as
shown in the comment\cite{Ericsson}. Therefore, the population
differences in Fig.1 are indeed induced by the non-Abelian gauge
potential. Moreover, we directly calculate the Schrodinger
equation $i\hbar \partial_t |\psi\rangle=H|\psi\rangle$ with the
Hamiltonian given by Eq.(1) in Ref.\cite{Ericsson}. We find that
the results are the same with those in Fig.1 when the parameters
are the same for the two different methods. It further confirms
that the population difference is indeed caused by the
non-vanished phase factor defined by the Wilson loop.

In summary, the main point that the authors used to defend in the
comment is that the induced gauge field is zero and thus the GP is
trivial. However, we have shown that this causality doesn't hold
for an SU(2) gauge field. Furthermore,  we have shown that the
non-Abelian GP in our proposal is non-trivial for its
non-vanishing closed loop integral and can be detected through the
induced significant population difference.

{\bf Acknowledge}: The authors thank Z. B. Li, D. W. Zhang and S.
L. Zhu for helpful discussions.  This work was supported by the
NSF of China(No. 11104085, No. 11125417, and No. 91121023 ), the
SKPBR of China(No. 2011CB922104) and PCSIRT. Du was aslo supported
by the SRFGS of SCNU.

\end{document}